\documentclass[aps,twocolumn,pra,preprintnumbers,showpacs,tightenlines]{revtex4}
\usepackage{amssymb}
\usepackage{amsmath}
\usepackage{graphicx}
\usepackage{epsfig}
\usepackage{subfigure}
\usepackage{amsfonts}
\usepackage{CJK}

\begin{document}

\title{Transferring arbitrary $d$-dimensional quantum states of a superconducting qudit in circuit QED}

\author{Tong Liu$^{1}$, Qi-Ping Su$^{1}$, Jin-Hu Yang$^{1}$, Yu Zhang$^{1}$, Shao-Jie Xiong$^{2}$, Jin-Ming Liu$^{2}$,}
\author{Chui-Ping Yang$^{1}$}
\email{yangcp@hznu.edu.cn}
\address{$^1$Department of Physics, Hangzhou Normal University, Hangzhou, Zhejiang 310036, China}
\address{$^2$ State Key Laboratory of Precision Spectroscopy, Department of Physics, East China Normal University, Shanghai
200062, China}
\date{\today}

\begin{abstract}

A qudit ($d$-level quantum systems) has a large Hilbert space and thus can be used to achieve many quantum information
and communication tasks. Here, we propose a method to transfer arbitrary $d$-dimensional quantum states (known or unknown) between two superconducting qudits coupled to a single cavity. The state transfer can be performed fast because of employing resonant interactions only. In addition, quantum states can be deterministically transferred without measurement. Numerical simulations show that high-fidelity transfer of quantum states between two superconducting transmon qudits ($d\leq5$) is feasible with current circuit QED technology. This proposal is quite general and can be applied to accomplish the same task with various superconducting qudits, quantum dots, or natural atoms coupled to a cavity or resonator.
\end{abstract}

\maketitle
\date{\today}

\begin{center}
\textbf{I. INTRODUCTION}
\end{center}

Many quantum information and communication tasks are usually based on qubits
(two-level quantum systems), but the use of qudits ($d$-level quantum
systems) can optimize some quantum computations \cite{s1,s2}, enhance the
security of quantum cryptography \cite{s3,s4}, realize bipartite
entanglement \cite{s5}, and simplify the implementation of quantum logic
gates \cite{s6,s7}. In addition, manipulation and measurement of a
superconducting phase qudit state or preparation and control of a transmon
qudit has been reported in experiments \cite{s8,s9}. Moreover, population transfer of
a three-level transmon qudit for $d=3$, via stimulated Raman adiabatic passage, has been
experimentally demonstrated recently \cite{K. S. Kumar}.

During the past years, superconducting qubits/qudits have been paid
intensive attention in quantum information and quantum computation due to
their significantly increased coherence times, controllability and
scalability \cite{s11,s12,s13,s17,s18,s19,s20,s21,s22}. Superconducting qubits/qudits based on Josephson junctions are mesoscopic
element circuits that behave like \textquotedblleft artificial atoms", whose
level spacings can be rapidly adjusted by varying external control
parameters (e.g., magnetic flux applied to the superconducting loop of a
superconducting phase, transmon, Xmon, or flux qubit/qudit; see, e.g., \cite%
{s19,s23,s24,s25}).

Circuit quantum electrodynamics (circuit QED) is analogue of cavity QED,
which has been considered as one of the most promising candidates for
quantum information processing (QIP) \cite{s12,s13,s26,s27,s28}.
The strong-coupling or ultrastrong-coupling regime with a
superconducitng qubit coupled to a microwave resonator has been
experimentally realized in circuit QED \cite{s29,Chiorescu,Forniaz,s30}.
Using superconducting qubits coupled to a single cavity or resonator, many
theoretical proposals have been presented for realizing quantum gates and
entanglement \cite{s26,s27,s28,Blais,s36,05Yang,Helmer,Bishop,10yang}.
Quantum effects and operations have been experimentally
demonstrated with superconducting qubits in circuit QED, including
demonstration of two-and three-qubit quantum gates \cite%
{s31,11Chow,s32,12Fedorov,s33}, realization of two-and three-qubit
entanglement \cite{s24,L. DiCarlo}, observation of Raman coherence effects
\cite{s34}, and suppression of dephasing by qubit motion \cite{s35}.
Moreover, a number of theoretical proposals have been proposed for realizing
quantum state transfer (QST) between two superconducting qubits through a
cavity \cite{s27,s36,s37,s38,s39,s40,s41,s42}. The QST between two
superconducting qubits has been experimentally demonstrated in circuit QED
\cite{s43,s44,s45,s46}.

The qudit-to-qudit QST plays a vital role in high-dimensional quantum
communication and QIP. Transfer of high-dimensional photon states through a
cavity array was previously proposed in \cite{Qin,15Liu}. In addition,
probabilistic transfer of high-dimensional quantum states between particles
via a spin chain has been studied \cite{14Bayat}. Moreover, a method has
been proposed for transferring quantum states between two superconducting
transmon qutrits via an adjustable inductive coupling \cite{Ghosh}, and an
approach has been presented for transferring quantum states between two
superconducting flux qutrits coupled to two resonators or cavities \cite{OL}%
. Here, qutrit refer to a three-level quantum system or a qudit for $d=3$.
Note that Refs. \cite{Ghosh,OL} only work for QST between two \textit{qutrits%
} and Ref. \cite{OL} requires the use of \textit{two} resonators or cavities
coupled to each qutrit.

Different from the previous works, we here propose a method to transfer
arbitrary $d$-dimensional quantum states (known or unknown) between two superconducting qudits coupled
to a single cavity. As shown below, this proposal has the following
advantages: (i) The experimental setup is very simple because only one cavity is used;
(ii) The speed of operation is fast due to using qudit-cavity and qudit-pulse resonant
interactions only; (iii) The QST can be deterministically achieved without measurement; and
(iv) The method can in principle be applied to transfer arbitrary $d$-dimensional quantum states between two $d$-level qudits for any positive integer $d$.
This proposal is quite general and can be applied to accomplish the same task with various superconducting qudits (e.g.,
superconducting transmon qudits, Xmon qudits, phase qudits, flux qudits),
quantum dots, or natural atoms coupled to a cavity or resonator.

This paper is organized as follows. In Sec.~II, we introduce the Hamiltonian
and time evolution of two qudits coupled to a cavity, as well as
the Hamiltonian and time evolution of a qudit driven by a classical pulse. In
Sec.~III, we show how to transfer arbitrary quantum states between two
superconducting qudits coupled to a cavity or resonator. In Sec.~IV, we discuss the
experimental feasibility of this proposal, by considering a setup of two
transmon qudits coupled to a 3D cavity and numerically calculating the
fidelity for the QST between two transom qudits for $d\leq5$. A
concluding summary is given in Sec.~V.

\begin{center}
\textbf{II. HAMILTONIAN AND TIME EVOLUTION }
\end{center}

Consider two qudits 1 and 2 coupled by a cavity. The cavity is resonant with
the transition between the two levels $|0\rangle $ and $|1\rangle $ of each
qudit. In the interaction picture, the Hamiltonian is given by~(in units of $%
\hbar =1$)
\begin{equation}
H_{I,1}=\sum_{j=1}^{2}g_{j}(a\sigma _{01,_{j}}^{+}+\text{h.c.}),
\end{equation}%
where $a$ is the photon annihilation operator for the cavity, the subscript $%
j$ represents qudit $j,$ $\sigma _{01,_{j}}^{+}=|1\rangle _{j}\langle 0|$,
and $g_{j}$ is the coupling constant between the cavity and the $|0\rangle
\leftrightarrow |1\rangle $ transition of qudit $j$ ($j=1,2$). For
simplicity, we set $g_{1}=g_{2}\equiv g$, which can be achieved by a prior
design of qudits or adjusting the position of each qudit located at the
cavity.

Under the Hamiltonian (1), one can obtain the following state evolutions:
\begin{eqnarray}
|0\rangle _{1}|0\rangle _{c}|0\rangle _{2} &\rightarrow &|0\rangle
_{1}|0\rangle _{c}|0\rangle _{2},  \notag \\
|1\rangle _{1}|0\rangle _{c}|0\rangle _{2} &\rightarrow &\frac{1}{2}(1+\cos
\sqrt{2}gt)|1\rangle _{1}|0\rangle _{c}|0\rangle _{2}  \notag \\
&-&\frac{\sqrt{2}}{2}i\sin (\sqrt{2}gt)|0\rangle _{1}|1\rangle _{c}|0\rangle
_{2}  \notag \\
&-&\frac{1}{2}(1-\cos \sqrt{2}gt)|0\rangle _{1}|0\rangle _{c}|1\rangle _{2},
\end{eqnarray}%
where $|0\rangle _{c}$ ($|1\rangle _{c}$) represents the vacuum (single photon)
state of the cavity and subscript 1 (2) represents qudit 1 (2).

We now consider applying a classical pulse to a qudit, which is resonant
with the transition between the level $\left\vert l-1\right\rangle $ and the
higher-energy level $\left\vert l\right\rangle $ of the qudit ($l=1,2,...,d-1
$). The Hamiltonian in the interaction picture is expressed as
\begin{equation}
H_{I,l}=\Omega (e^{i\phi }\left\vert l-1\right\rangle \left\langle
l\right\vert +\text{h.c.}),
\end{equation}%
where $\Omega $ and $\phi $ are the Rabi frequency and the initial phase of
the pulse. One can obtain the following rotations under the Hamiltonian~(3),
\begin{eqnarray}
|l-1\rangle  &\rightarrow &\cos (\Omega t)|l-1\rangle -ie^{-i\phi }\sin
(\Omega t)|l\rangle ,  \notag \\
|l\rangle  &\rightarrow &\cos (\Omega t)|l\rangle -ie^{i\phi }\sin (\Omega
t)|l-1\rangle .
\end{eqnarray}%
The results given in Eqs. (2) and (4) will be employed for implementing QST
between two qudits, which is described in the next section.

\begin{figure}[tbp]
\begin{center}
\includegraphics[bb=80 635 512 767, width=8.5 cm, clip]{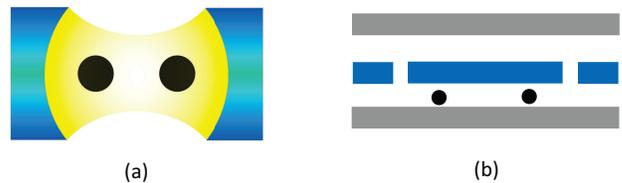} \vspace*{%
-0.08in}
\end{center}
\caption{(color online) (a) Setup for two superconducting qudits embedded in
a 3D cavity. (b) Setup for two superconducting qudits coupled to a 1D
transmission line resonator. A dark dot represents a superconducting qudit.}
\label{fig:1}
\end{figure}

\begin{figure}[tbp]
\begin{center}
\includegraphics[bb=68 42 568 778, width=8.5 cm, clip]{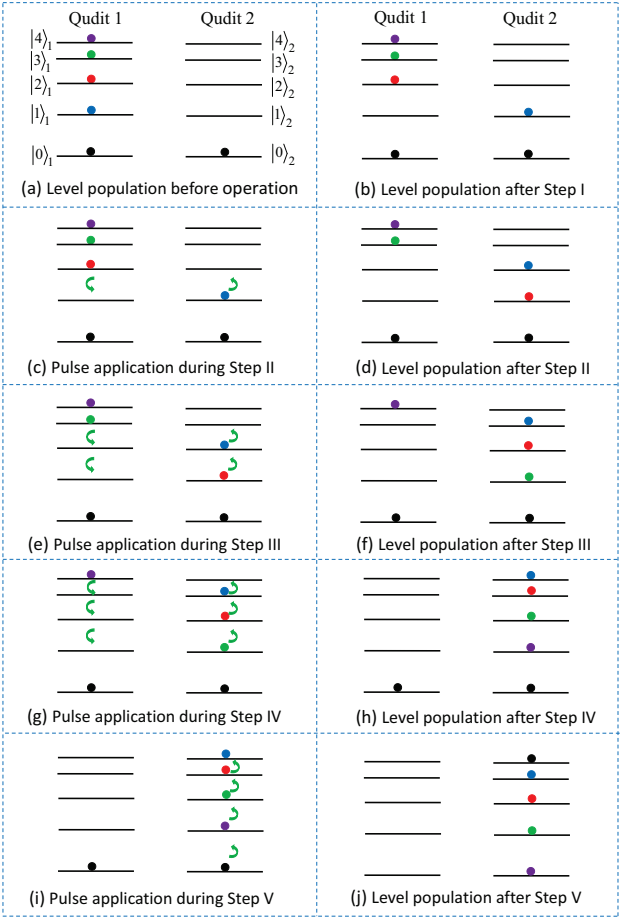} \vspace*{%
-0.08in}
\end{center}
\caption{(color online) The color circles indicate the occupied energy
levels. Each green arrow represents a classical pulse, which is resonant
with the transition between the two neighbor levels close to each green
arrow. In (e) and (g), the sequence for applying the pulses is from top to
bottom, and the lower pulses are turned on after the upper pulses are
switched off. In (i), the sequence for applying the pulses is from bottom to
top, and the upper pulses are turned on after the lower pulses are switched
off. For the details on the applied pulses, see the descriptions given in
the text. Note that in (a)-(j), the left levels are for qudit 1 while the
right levels are for qudit 2. For simplicity, we here consider the case that the spacings between
adjacent levels become narrow as the levels move up, which is actually unnecessary.}
\label{fig:2}
\end{figure}

\begin{center}
\textbf{III. QUANTUM STATE TRANSFER BETWEEN TWO SUPERCONDUCTING QUDITS}
\end{center}

Our system, shown in Fig.~1, consists of two superconducting qudits 1 and 2
embedded in a 3D microwave cavity or coupled to a 1D resonator. As an
example, we will explicitly show how to transfer quantum states between two
qudits for $d\leq 5$. We then give a brief discussion on how to extend the
method to transfer arbitrary $d$-dimensional quantum states between two $d$-level qudits for any
positive integer $d$.

Without loss of generality, we here consider qudits with a ladder-type level
structure formed by $d$ levels $\left\vert 0\right\rangle ,$ $\left\vert
1\right\rangle ,$ $\left\vert 2\right\rangle ,...$ and $\left\vert
d-1\right\rangle $ (Fig.2 for $d=5$)$.$ For a ladder-type level structure,
the transition between adjacent levels is allowed but the transition between
non-adjacent levels is forbidden or very weak. Note that this ladder-type level
structure is available in superconducting transmon qudits, Xmon qudits,
phase qudits, quantum dots, or nature atoms. In the following, the transition frequency
between two adjacent levels $\left\vert l-1\right\rangle $ and $\left\vert
l\right\rangle $ of each qudit is labeled as $\omega _{(l-1)l}$ ($l=1,2,...,d-1
$). The initial phase, duration, and frequency of the pulses are denoted as $%
\{\phi ,t,\omega \}$. For simplicity, we set the same Rabi frequency $\Omega
$ for each pulse, which can be readily achieved by adjusting the pulse
intensity.

\begin{center}
\textbf{A. Case for $d=5$}
\end{center}

The five levels of qudits are labeled as $\left\vert 0\right\rangle ,$ $%
\left\vert 1\right\rangle ,$ $\left\vert 2\right\rangle ,$ $\left\vert
3\right\rangle ,$ and $\left\vert 4\right\rangle $ (Fig.2). Assume that
qudit 1 is initially in an arbitrary quantum state $\sum_{l=0}^{4}c_{l}|l%
\rangle _{1}$ (known or unknown) with level populations illustrated in Fig. 2(a), qudit 2 is
initially in the ground state $\left\vert 0\right\rangle _{2}$, and the
cavity is initially in the vacuum state $\left\vert 0\right\rangle _{c}$.
Here and below, $c_{l}$ is a normalized coefficient.

To begin with, the level spacings of the qudits need to be adjusted to have
the cavity resonant with the $|0\rangle \leftrightarrow |1\rangle $
transition of each qudit. The procedure for implementing the QST from qudit
1 to qudit 2 is described as follows:

Step I. Let the cavity resonant with the $|0\rangle \leftrightarrow
|1\rangle $ transition of each qudit described by Hamiltonian (1). According
to Eq.~(2), after an interaction time $t_{1}=\pi /(\sqrt{2}g)$, one has the
state transformation
\begin{equation}
\left\vert 1\right\rangle _{1}\left\vert 0\right\rangle _{c}\left\vert
0\right\rangle _{2}\rightarrow -\left\vert 0\right\rangle _{1}\left\vert
0\right\rangle _{c}\left\vert 1\right\rangle _{2},
\end{equation}%
which shows that the cavity remains in the vacuum state after the
qudit-cavity interaction. Thus, the initial state $\sum_{l=0}^{4}c_{l}|l%
\rangle _{1}\otimes \left\vert 0\right\rangle _{2}$ of the two qudits
becomes
\begin{equation}
\left( c_{0}\left\vert 0\right\rangle _{1}+c_{2}\left\vert 2\right\rangle
_{1}+c_{3}\left\vert 3\right\rangle _{1}+c_{4}\left\vert 4\right\rangle
_{1}\right) |0\rangle _{2}-c_{1}\left\vert 0\right\rangle _{1}\left\vert
1\right\rangle _{2}.
\end{equation}%
Eq. (6) shows that the population of the level $\left\vert 1\right\rangle $
of qudit 1 is transferred onto the level $\left\vert 1\right\rangle $ of
qudit 2 [Fig. 2(b)].

Step II. Apply a pulse of $\{\pi /2,\pi /2\Omega ,\omega _{12}\}$ to qudit 1
while a pulse of $\{-\pi /2,\pi /2\Omega ,\omega _{12}\}$ to qudit 2 [Fig.
2(c)]. According to Eq.~(4), the pulses lead to $\left\vert 2\right\rangle
_{1}\rightarrow \left\vert 1\right\rangle _{1}$ and $\left\vert
1\right\rangle _{2}\rightarrow \left\vert 2\right\rangle _{2}$. Thus, the
state (6) becomes
\begin{equation}
\left( c_{0}\left\vert 0\right\rangle _{1}+c_{2}\left\vert 1\right\rangle
_{1}+c_{3}\left\vert 3\right\rangle _{1}+c_{4}\left\vert 4\right\rangle
_{1}\right) |0\rangle _{2}-c_{1}\left\vert 0\right\rangle _{1}\left\vert
2\right\rangle _{2}.
\end{equation}%
For $\Omega \gg g$, the interaction between the cavity and the qudits can be
neglected during the pulse. Now let the cavity resonant with the $|0\rangle
\leftrightarrow |1\rangle $ transition of each qudit for an interaction time
$t_{2}=\pi /(\sqrt{2}g)$, to obtain the state transformation (5). Hence, the
state (7) becomes
\begin{equation}
\left( c_{0}\left\vert 0\right\rangle _{1}+c_{3}\left\vert 3\right\rangle
_{1}+c_{4}\left\vert 4\right\rangle _{1}\right) |0\rangle _{2}+\left(
-c_{2}\left\vert 1\right\rangle _{2}-c_{1}\left\vert 2\right\rangle
_{2}\right) \left\vert 0\right\rangle _{1},
\end{equation}%
which shows that the populations for the levels $\left\vert 1\right\rangle $
and $\left\vert 2\right\rangle $ of qudit 1 are transferred onto the levels $%
\left\vert 2\right\rangle $ and $\left\vert 1\right\rangle $ of qudit 2,
respectively [Fig.~2(d)].

Step III. Apply a pulse of $\{\pi /2,\pi /2\Omega ,\omega _{23}\}$ and then
a pulse of $\{\pi /2,\pi /2\Omega ,\omega _{12}\}$ to qudit 1, while a pulse
of $\{-\pi /2,\pi /2\Omega ,\omega _{23}\}$ and then a pulse of $\{-\pi
/2,\pi /2\Omega ,\omega _{12}\}$ to qudit 2 [Fig. 2(e)]. The pulses result
in the transformations $\left\vert 3\right\rangle _{1}\rightarrow \left\vert
1\right\rangle _{1}$ (via $\left\vert 3\right\rangle _{1}\rightarrow
\left\vert 2\right\rangle _{1}\rightarrow \left\vert 1\right\rangle _{1}$), $%
\left\vert 2\right\rangle _{2}\rightarrow \left\vert 3\right\rangle _{2}$
and $\left\vert 1\right\rangle _{2}\rightarrow \left\vert 2\right\rangle
_{2}.$ Thus, the state (8) becomes
\begin{equation}
\left( c_{0}\left\vert 0\right\rangle _{1}+c_{3}\left\vert 1\right\rangle
_{1}+c_{4}\left\vert 4\right\rangle _{1}\right) |0\rangle _{2}+\left(
-c_{2}\left\vert 2\right\rangle _{2}-c_{1}\left\vert 3\right\rangle
_{2}\right) \left\vert 0\right\rangle _{1}.
\end{equation}%
Let the cavity resonant with the $|0\rangle \leftrightarrow |1\rangle $
transition of each qudit for an interaction time $t_{3}=\pi /(\sqrt{2}g)$,
to achieve the state transformation (5). Thus, the state (9) becomes
\begin{equation}
\left( c_{0}\left\vert 0\right\rangle _{1}+c_{4}\left\vert 4\right\rangle
_{1}\right) |0\rangle _{2}+\left( -c_{3}\left\vert 1\right\rangle
_{2}-c_{2}\left\vert 2\right\rangle _{2}-c_{1}\left\vert 3\right\rangle
_{2}\right) \left\vert 0\right\rangle _{1},
\end{equation}%
which shows that the populations for the levels $\left\vert 1\right\rangle ,$
$\left\vert 2\right\rangle ,$ and $\left\vert 3\right\rangle $ of qudit 1
are transferred onto the levels $\left\vert 3\right\rangle ,$ $\left\vert
2\right\rangle ,$ and $\left\vert 1\right\rangle $ of qudit 2, respectively
[Fig.~2(f)].

Step IV. Apply pulses of $\{\pi /2,\pi /2\Omega ,\omega _{34}\}$, $\{\pi
/2,\pi /2\Omega ,\omega _{23}\}$ and then $\{\pi /2,\pi /2\Omega ,\omega
_{12}\}$ to qudit 1 while pulses of $\{-\pi /2,\pi /2\Omega ,\omega _{34}\}$%
, $\{-\pi /2,\pi /2\Omega ,\omega _{23}\}$ and then $\{-\pi /2,\pi /2\Omega
,\omega _{12}\}$ to qudit 2 [Fig. 2(g)], which leads to the transformations $%
\left\vert 4\right\rangle _{1}\rightarrow \left\vert 1\right\rangle _{1}$
(via $\left\vert 4\right\rangle _{1}\rightarrow \left\vert 3\right\rangle
_{1}\rightarrow \left\vert 2\right\rangle _{1}\rightarrow \left\vert
1\right\rangle _{1}$), $\left\vert 3\right\rangle _{2}\rightarrow \left\vert
4\right\rangle _{2},$ $\left\vert 2\right\rangle _{2}\rightarrow \left\vert
3\right\rangle _{2}$ and $\left\vert 1\right\rangle _{2}\rightarrow
\left\vert 2\right\rangle _{2}.$ Hence, the state (10) becomes
\begin{equation}
\left( c_{0}\left\vert 0\right\rangle _{1}+c_{4}\left\vert 1\right\rangle
_{1}\right) |0\rangle _{2}+\left( -c_{3}\left\vert 2\right\rangle
_{2}-c_{2}\left\vert 3\right\rangle _{2}-c_{1}\left\vert 4\right\rangle
_{2}\right) \left\vert 0\right\rangle _{1}.
\end{equation}%
Let the cavity resonant with the $|0\rangle \leftrightarrow |1\rangle $
transition of each qudit for an interaction time $t_{4}=\pi /(\sqrt{2}g)$,
to have the state transformation (5). Thus, the state (11) changes
\begin{equation}
\left( c_{0}\left\vert 0\right\rangle _{2}-c_{4}\left\vert 1\right\rangle
_{2}-c_{3}\left\vert 2\right\rangle _{2}-c_{2}\left\vert 3\right\rangle
_{2}-c_{1}\left\vert 4\right\rangle _{2}\right) \left\vert 0\right\rangle
_{1},
\end{equation}%
which shows that the populations for the levels $\left\vert 1\right\rangle
_{1},\left\vert 2\right\rangle _{1},\left\vert 3\right\rangle _{1},$ and $%
\left\vert 4\right\rangle _{1}$ of qudit 1 have been transferred onto the
levels $\left\vert 4\right\rangle _{2},\left\vert 3\right\rangle
_{2},\left\vert 2\right\rangle _{2},$ and $\left\vert 1\right\rangle _{2}$
of qudit 2, respectively [Fig.~2(h)]. After this step of operation, to
maintain the state (12), the level spacings of the qudits need to be
adjusted so that the qudits are decoupled from the cavity.

Step V. By sequentially applying pulses of $\{-\pi /2,\pi /2\Omega ,\omega
_{01}\}$, $\{-\pi /2,\pi /2\Omega ,\omega _{12}\},$ $\{-\pi /2,\pi /2\Omega $%
, $\omega _{23}\},$ and then $\{-\pi /2,\pi /2\Omega ,\omega _{34}\}$ to
qudit 2 [Fig. 2(i)], we obtain the state transformations $\left\vert
0\right\rangle _{2}\rightarrow \left\vert 4\right\rangle _{2}$ (via $%
\left\vert 0\right\rangle _{2}\rightarrow \left\vert 1\right\rangle
_{2}\rightarrow \left\vert 2\right\rangle _{2}\rightarrow \left\vert
3\right\rangle _{2}\rightarrow \left\vert 4\right\rangle _{2}$)$,$ $%
\left\vert 1\right\rangle _{2}\rightarrow -\left\vert 0\right\rangle _{2},$ $%
\left\vert 2\right\rangle _{2}\rightarrow -\left\vert 1\right\rangle _{2},$ $%
\left\vert 3\right\rangle _{2}\rightarrow -\left\vert 2\right\rangle _{2},$
and $\left\vert 4\right\rangle _{2}\rightarrow -\left\vert 3\right\rangle
_{2}.$ Hence, the state (12) becomes
\begin{equation}
\left( c_{4}\left\vert 0\right\rangle _{2}+c_{3}\left\vert 1\right\rangle
_{2}+c_{2}\left\vert 2\right\rangle _{2}+c_{1}\left\vert 3\right\rangle
_{2}+c_{0}\left\vert 4\right\rangle _{2}\right) \left\vert 0\right\rangle
_{1}.
\end{equation}%
The result (13) shows that an arbitrary quantum state $\sum_{l=0}^{4}c_{l}|l%
\rangle _{1}$ of qudit 1 has been transferred onto qudit 2 via the population
transfer from the five levels \{$\left\vert 0\right\rangle _{1},\left\vert
1\right\rangle _{1},\left\vert 2\right\rangle _{1},\left\vert 3\right\rangle
_{1},\left\vert 4\right\rangle _{1}\}$ of qudit 1 to the five levels \{$%
\left\vert 4\right\rangle _{2},\left\vert 3\right\rangle _{2},\left\vert
2\right\rangle _{2},\left\vert 1\right\rangle _{2},\left\vert 0\right\rangle
_{2}\}$ of qudit 2, respectively [Fig.~2(j)].

\begin{center}
\textbf{B. Case for $d=4$ and $d=3$}
\end{center}

From the above description, it can be found that by performing the
operations of steps I, II, and III above, and then by sequentially applying
pulses of $\{-\pi /2,\pi /2\Omega ,\omega _{01}\}$, $\{-\pi /2,\pi /2\Omega
,\omega _{12}\},$ and $\{-\pi /2,\pi /2\Omega $, $\omega _{23}\}$ to qudit
2, we can obtain the state transformation $\sum_{l=0}^{3}c_{l}|l\rangle
_{1}\otimes |0\rangle _{2}\rightarrow |0\rangle _{1}\otimes \left(
c_{0}\left\vert 3\right\rangle _{2}+c_{1}\left\vert 2\right\rangle
_{1}+c_{2}\left\vert 1\right\rangle _{2}+c_{3}\left\vert 0\right\rangle
_{2}\right) ,$ which implies that the QST for $d=4$ is implemented, i.e., an
arbitrary quantum state of qudit 1 is transferred onto
qudit 2 via the population transfer from the four levels \{$\left\vert
0\right\rangle _{1},\left\vert 1\right\rangle _{1},\left\vert 2\right\rangle
_{1},\left\vert 3\right\rangle _{1}\}$ of qudit 1 to the four levels \{$%
\left\vert 3\right\rangle _{2},\left\vert 2\right\rangle _{2},\left\vert
1\right\rangle _{2},\left\vert 0\right\rangle _{2}\}$ of qudit 2,
respectively.

By performing the operations of steps I and II above, followed by applying
pulses of $\{-\pi /2,\pi /2\Omega ,\omega _{01}\}$ and then $\{-\pi /2,\pi
/2\Omega ,\omega _{12}\}$ to qudit 2, the state transformation $%
\sum_{l=0}^{2}c_{l}|l\rangle _{1}\otimes |0\rangle _{2}\rightarrow |0\rangle
_{1}\otimes \left( c_{0}\left\vert 2\right\rangle _{2}+c_{1}\left\vert
1\right\rangle _{1}+c_{2}\left\vert 0\right\rangle _{2}\right) $ can be
achieved, which shows that the QST for $d=3$ (i.e., the QST between two
qutrits)\ is realized, i.e., an arbitrary quantum state of
qudit 1 is transferred onto qudit 2 via transferring the populations of the
three levels \{$\left\vert 0\right\rangle _{1},\left\vert 1\right\rangle
_{1},\left\vert 2\right\rangle _{1}\}$ of qudit 1 to the three levels \{$%
\left\vert 2\right\rangle _{2},\left\vert 1\right\rangle _{2},\left\vert
0\right\rangle _{2}\}$ of qudit 2, respectively.

\begin{center}
\textbf{C. Case for any positive integer $d$}
\end{center}

By examining the operations introduced in subsection IIIA (i.e., QST for $d=5$),
one can easily find that an arbitrary $d$-dimensional quantum state can be transferred
between two $d$-level qudits for any positive integer $d$, through the following $d$ operational steps.
The first operational step is the same as that
described in step \uppercase\expandafter{\romannumeral 1} above. For the $%
l^{th}$ operational step ($1<l<d$), $l-1$ pulses of $\{\pi /2,\pi /2\Omega
,\omega _{(l-1)l}\},\cdots ,\{\pi /2,\pi /2\Omega ,$$\omega _{23}\},$ and $%
\{\pi /2,\pi /2\Omega ,\omega _{12}\}$ should be applied to qudit 1 in turn
(from left to right), while other $l-1$ pulses of $\{-\pi /2,\pi /2\Omega ,$$%
\omega _{(l-1)l}\},$$\cdots ,\{-\pi /2,$ $\pi /2\Omega ,$$\omega _{23}\},$
and $\{-\pi /2,\pi /2\Omega ,\omega _{12}\}$ should be applied to qudit 2 in
sequence (from left to right), followed by each qudit simultaneously
resonantly interacting with the cavity for an interaction time $t=\pi /(%
\sqrt{2}g)$. One can easily check that after the first $d-1$ steps of
operation, the following state transformation can be obtained $%
\sum_{l=0}^{d-1}c_{l}|l\rangle _{1}\otimes |0\rangle _{2}\rightarrow
|0\rangle _{1}\otimes \left( c_{0}\left\vert 0\right\rangle
_{2}-\sum_{l=1}^{d-1}c_{l}|d-l\rangle _{2}\right) ,$ which can further turn
into
\begin{equation}
\sum_{l=0}^{d-1}c_{l}|l\rangle _{1}\otimes |0\rangle _{2}\rightarrow
|0\rangle _{1}\otimes \left( c_{0}\left\vert d-1\right\rangle
_{2}+c_{1}\left\vert d-2\right\rangle _{1}+\cdot \cdot \cdot
+c_{d-1}\left\vert 0\right\rangle _{2}\right) ,
\end{equation}%
by sequentially applying pulses of $\{-\pi /2,\pi /2\Omega ,\omega _{01}\}$,
$\{-\pi /2,\pi /2\Omega ,\omega _{12}\},...,$ and then $\{-\pi /2,\pi
/2\Omega ,\omega _{\left( d-2\right) \left( d-1\right) }\}$ to qudit 2
(i.e., the last step of operation). The result (14) implies that an
arbitrary $d$-dimensional quantum state of qudit 1 (known or unknown) has been transferred onto qudit 2 through
the population transfer from the $d$ levels \{$\left\vert 0\right\rangle
_{1},\left\vert 1\right\rangle _{1},\left\vert 2\right\rangle
_{1},...,\left\vert d-2\right\rangle _{1},\left\vert d-1\right\rangle _{1}$%
\} of qudit 1 to the $d$ levels \{$\left\vert d-1\right\rangle
_{2},\left\vert d-2\right\rangle _{2},...,\left\vert 2\right\rangle
_{2},\left\vert 1\right\rangle _{2},\left\vert 0\right\rangle _{2}$\} of
qudit 2, respectively. It is straightforward to show that the state $%
c_{0}\left\vert d-1\right\rangle _{2}+c_{1}\left\vert d-2\right\rangle
_{1}+\cdot \cdot \cdot +c_{d-1}\left\vert 0\right\rangle _{2}$ can be
converted into the state $\sum_{l=0}^{d-1}c_{l}|l\rangle _{2}$ by applying a
series of pulses on qudit 2. Note that the number of pulses required and the sequence of
pulses depend on the level structures of the qudits.

From the description given above, one can see that the procedure described
above for QST does not employ the coupling of the cavity/pulse with the
transition between non-adjacent levels. In this sense, this proposal can be
applied to implement QST for qudits with other types of level structures
(not just limited to the ladder-type level structure).  For qudits with
other types of level structure, the effect of the unwanted coupling of the cavity/pulse
with the transition between non-adjacent levels can be made to be negligibly
small as long as the cavity/pulse is highly detuned from the transition
between non-adjacent levels (which can be achieved by adjusting the level
spacings of qudits). As mentioned previously, the level spacings of
superconducting qudits can be rapidly adjusted by varying external control
parameters \cite{s19,s23,s24,s25}. In addition, the level spacings of atoms/quantum dots
can be adjusted by changing the voltage on the electrodes around each
atom/quantum dot \cite{2000Pradhan}.

\begin{center}
\textbf{IV. POSSIBLE EXPERIMENTAL IMPLEMENTATION}
\end{center}

So far we have considered a general type of superconducting qudit. For an
experimental implementation, let us now consider a setup of two
superconducting transmon qudits embedded in a 3D cavity. This architecture
is feasible in the state-of-the-art superconducting setup as demonstrated
recently in [9]. For simplicity, we consider QST between the two transmon
qudits 1 and 2 for $d\leq 5.$ As an example, suppose that the state of qudit
1 to be transferred is: (i) $\frac{1}{\sqrt{3}}\sum_{l=0}^{2}|l\rangle _{1}$
for $d=3,$ (ii) $\frac{1}{2}\sum_{l=0}^{3}|l\rangle _{1}$ for $d=4,$ and
(iii) $\frac{1}{\sqrt{5}}\sum_{l=0}^{4}|l\rangle _{1}$ for $d=5.$

We take into account the influence of the unwanted coupling of the cavity
with the $\left\vert 1\right\rangle \leftrightarrow \left\vert
2\right\rangle $ transition. The Hamiltonian $H_{I,1}$ is thus modified as
\begin{equation}
H_{I,1}^{\prime }=H_{I,1}+\varepsilon _{1},
\end{equation}%
where $\varepsilon _{1}$ describes the unwanted off-resonant coupling
between the cavity and the $\left\vert 1\right\rangle \leftrightarrow
\left\vert 2\right\rangle $ transition of each qudit, which is given by $%
\varepsilon _{1}=\sum_{j=1}^{2}\widetilde{g}_{j}(a\sigma _{12,_{j}}^{+}+%
\text{h.c.})$ where $\sigma _{12,_{j}}^{+}=|2\rangle _{j}\langle 1|.$ For a
transmon qudit, one has $\widetilde{g_{j}}\sim \sqrt{2}g_{j}$ \cite{s47}.

We also consider the effect of the unwanted couplings of the pulse with the two adjacent $%
\left\vert l-2\right\rangle \leftrightarrow \left\vert l-1\right\rangle $
and $\left\vert l\right\rangle \leftrightarrow \left\vert l+1\right\rangle $
transitions, when the pulse is resonant with the $\left\vert
l-1\right\rangle \leftrightarrow \left\vert l\right\rangle $ transition of
each qudit. Here and below, $l\in \{1,2,3,4\}$ for $d=5,$ $l\in \{1,2,3\}$
for $d=4,$ and $l\in \{1,2\}$ for $d=3.$ After this consideration, the
Hamiltonian $H_{I,l}$ is modified as
\begin{equation}
H_{I,l}^{\prime }=H_{I,l}+H_{I,1}^{\prime }+\varepsilon _{l},
\end{equation}%
where $\varepsilon _{l}$ describes the unwanted off-resonant couplings of
the pulse with the $\left\vert l-2\right\rangle \leftrightarrow \left\vert
l-1\right\rangle $ and $\left\vert l\right\rangle \leftrightarrow \left\vert
l+1\right\rangle $ transitions of each qudit, during the pulse resonant with
the $\left\vert l-1\right\rangle \leftrightarrow \left\vert l\right\rangle $
transition of each qudit. Here, $\varepsilon _{l}$ is given by $\left(
\Omega /\sqrt{2}\right) (e^{i\phi }\left\vert l-2\right\rangle \left\langle
l-1\right\vert +$h.c.$)$ $+\sqrt{2}\Omega (e^{i\phi }\left\vert
l\right\rangle \left\langle l+1\right\vert +$h.c.$)$ \cite{s47}. Note that
the effect of the cavity-qudit interaction during the pulse application is
also considered here, which is described by the $H_{I,1}^{\prime }.$

For a transmon qudit, the transition between non-adjacent levels
is forbidden or very weak \cite{s47}. Thus, the couplings of
the cavity/pulses with the transitions between non-adjacent levels can be neglected.
In addition, the spacings between adjacent levels for a transmon qudit
become narrow as the levels move up (Fig. 2). Therefore, the detunings between the
cavity frequency and the transition frequencies for adjacent levels (e.g., levels $\left\vert 1\right\rangle$ and $\left\vert 2\right\rangle$,
levels $\left\vert 2\right\rangle$ and $\left\vert 3\right\rangle $, levels $\left\vert 3\right\rangle$ and
$\left\vert 4\right\rangle $, etc.) increase when the levels go up. As
a result, when compared with the coupling effect of the cavity with
the $\left\vert 1\right\rangle \leftrightarrow \left\vert 2\right\rangle $ transition, the coupling effect of
the cavity with the transitions for other adjacent levels is negligibly small, which is
thus not considered in the numerical simulation for simplicity. For similar reasons, when the pulse is resonant
with the $\left\vert l-1\right\rangle \leftrightarrow \left\vert l\right\rangle $ transition of
each qudit, the coupling effect of the pulses with the transitions between other adjacent levels is weak and thus
we only consider the effect of the coupling of the pulse with the two adjacent $%
\left\vert l-2\right\rangle \leftrightarrow \left\vert l-1\right\rangle $
and $\left\vert l\right\rangle \leftrightarrow \left\vert l+1\right\rangle $
transitions.

When the dissipation and dephasing are included, the dynamics of the lossy
system is determined by the following master equation
\begin{eqnarray}
\frac{d\rho }{dt} &=&-i\left[ H^{\prime },\rho \right] +\kappa \mathcal{L}%
\left[ a\right]   \notag \\
&+&\sum_{j=1}^{2}\sum\limits_{l=1}^{d-1}\left\{ \gamma _{\left( l-1\right)
l,_{j}}\mathcal{L}\left[ \sigma _{\left( l-1\right) l,_{j}}^{-}\right]
\right\}   \notag \\
&+&\sum_{j=1}^{2}\sum_{l=1}^{d-1}\left\{ \gamma _{\varphi l,_{j}}\left(
\sigma _{ll,_{j}}\rho \sigma _{ll,_{j}}-\sigma _{ll,_{j}}\rho /2-\rho \sigma
_{ll,_{j}}/2\right) \right\} ,\ \ \ \
\end{eqnarray}%
where $d\in \{3,4,5\},$ $H^{\prime }$ is the modified Hamiltonian $%
H_{I,1}^{\prime }$ or $H_{I,l}^{\prime }$ given above, $\sigma _{\left(
l-1\right) l,_{j}}^{-}=\left\vert l-1\right\rangle _{j}\left\langle
l\right\vert ,\sigma _{ll,_{j}}=\left\vert l\right\rangle _{j}\left\langle
l\right\vert ,$ and $\mathcal{L}\left[ \Lambda \right] =\Lambda \rho \Lambda
^{+}-\Lambda ^{+}\Lambda \rho /2-\rho \Lambda ^{+}\Lambda /2$ with $\Lambda
=a,\sigma _{\left( l-1\right) l,_{j}}^{-}.$ Here, $\kappa $ is the photon
decay rate of the cavity. In addition, $\gamma _{\left( l-1\right) l,_{j}}$
is the energy relaxation rate of the level $\left\vert l\right\rangle $ for
the decay path $\left\vert l\right\rangle \rightarrow \left\vert
l-1\right\rangle $ and $\gamma _{\varphi l,_{j}}$ is the dephasing rate of
the level $\left\vert l\right\rangle $ of qudit $j$ ($j=1,2$).

The fidelity of the operation is given by $\mathcal{F}=\sqrt{\left\langle
\psi _{id}\right\vert {\rho }\left\vert \psi _{id}\right\rangle }$, where $%
\left\vert \psi _{id}\right\rangle $ is the output state of an ideal system
(i.e., without dissipation and dephasing considered), which is given by: (i)
$\left\vert \psi _{id}\right\rangle =|0\rangle _{1}|0\rangle _{c}\otimes (1/%
\sqrt{3})\sum_{l=0}^{2}|l\rangle _{2}$ for $d=3$, (ii) $\left\vert \psi
_{id}\right\rangle =|0\rangle _{1}|0\rangle _{c}\otimes
(1/2)\sum_{l=0}^{3}|l\rangle _{2}$ for $d=4$, and (iii) $\left\vert \psi
_{id}\right\rangle =|0\rangle _{1}|0\rangle _{c}\otimes (1/\sqrt{5}%
)\sum_{l=0}^{4}|l\rangle _{2}$ for $d=5$. Note that $\rho $ is the final
density operator of the system when the operation is performed in a
realistic situation.

\begin{figure}[tbp]
\begin{center}
\includegraphics[bb=-1 292 595 784, width=8.5 cm, clip]{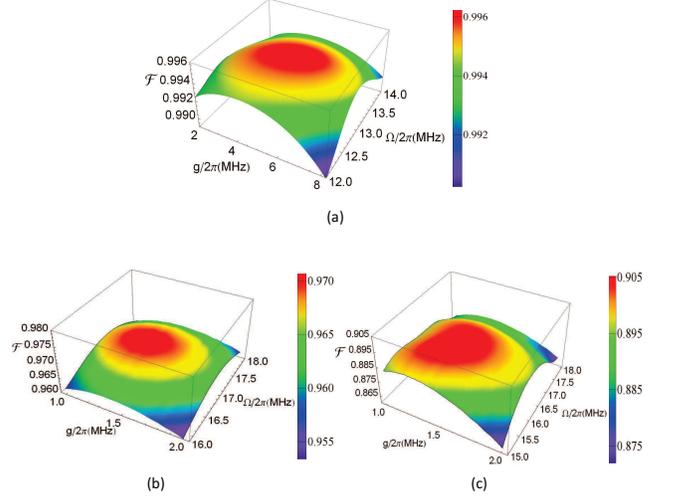} \vspace*{%
-0.08in}
\end{center}
\caption{(color online) Fidelity for the quantum state transfer for $g/2%
\protect\pi $ and $\Omega /2\protect\pi $. (a) Fidelity for $d=3$. (b)
Fidelity for $d=4$. (c) Fidelity for $d=5$.}
\label{fig:3}
\end{figure}

Without loss of generality, consider identical transmon qudits. In this
case, the decoherence rates are the same for each qudit and thus the
subscript $j$ involved in the decoherence rates above can be omitted.
According to \cite{s9}, we choose $(\omega _{01}-\omega _{12})/2\pi =275$
MHz, $(\omega _{12}-\omega _{23})/2\pi =309$ MHz, and $(\omega _{23}-\omega
_{34})/2\pi =358$ MHz. The decoherence times for the qudits and the cavity,
used in the numerical calculation, are as follows: $\gamma _{01}^{-1}=84$ $%
\mu $s, $\gamma _{12}^{-1}$ = 41 $\mu $s, $\gamma _{23}^{-1}$ = 30 $\mu $s, $%
\gamma _{34}^{-1}$ = 22 $\mu $s, $\gamma _{\varphi 1}^{-1}=72$ $\mu $s, $%
\gamma _{\varphi 2}^{-1}$ = 32 $\mu $s, $\gamma _{\varphi 3}^{-1}$ = 12 $\mu
$s, $\gamma _{\varphi 4}^{-1}$ = 2 $\mu $s, and $\kappa ^{-1}=15$ $\mu $s.
The decoherence times of transmon qudits considered here are realistic
because they are from the recent experimental report in \cite{s9}. In a
realistic situation, it may be a challenge to obtain exact identical
qudit-resonator couplings. Therefore, we consider inhomogeneous
qudit-resonator couplings, e.g., $g_{1}=g$ and $g_{2}=0.95g$.

We numerically calculate the fidelity of the entire operation based on the
master equation. Figure~3(a,b,c) shows the fidelity versus $%
g/2\pi $ and $\Omega /2\pi $ for QST between two qudits for $d=3$, $d=4$,
and $d=5$, respectively. From Fig.~3(a), one can see that for $g/2\pi $ $\in
\lbrack 2,8]$ MHz and $\Omega /2\pi $ $\in \lbrack 12,14]$ MHz, the fidelity
can be greater than $98.8\%$ for $d=3$. When $g/2\pi =$ 5.4 MHz and $\Omega
/2\pi $= 12.8 MHz, the fidelity value is the optimum with a value of $\sim
99.6\%$ for $d=3$. As shown in Fig.~3(b), the value of the fidelity has a
slow decline for $d=4$. In Fig.~3(b) the optimal value for $\mathcal{F}$$%
\sim 96.96\%$ is obtained for $g/2\pi =$ 1.35 MHz and for $\Omega /2\pi $=
17.00 MHz. While $\mathcal{F}$ drastically decreases for $d=5$, a high
fidelity $\sim 90.32\%$ is attainable with $g/2\pi =$ 1.45 MHz and $\Omega
/2\pi $= 16.00 MHz [see Fig.~3(c)]. Note that the above values of the $g$
and $\Omega $ are readily available in experiments \cite{s48,s49,09M. Baur,14F. Yoshihara}.

For a cavity with frequency $\omega _{c}/2\pi =4.97$ GHz and dissipation
time $\kappa ^{-1}$ used in the numerical simulation, the quality factor of
the cavity is $Q\sim 4.7\times 10^{5}$. Note that three-dimensional cavities
with a loaded quality factor $Q>10^{6}$ have been implemented in experiments
\cite{s48,16M. Reagor}.

\begin{center}
\textbf{V. CONCLUSION}
\end{center}

We have presented a method to deterministically transfer arbitrary $d$%
-dimensional quantum states (known or unknown) between two superconducting
qudits, which are coupled to a single cavity or resonator. As shown above,
only a single cavity or resonator is needed, thus the experimental setup is very simple.
The state transfer can be performed fast because of employing resonant interactions only.
In addition, no measurement is required. Numerical simulation shows that
high-fidelity transfer of quantum states between two transmon qudits for ($d\leq 5$) is
feasible with current circuit-QED technology. This proposal is general and can be applied
to accomplish the same task with various superconducting qudits, quantum dots, or natural
atoms coupled to a cavity or resonator. We remark that the number of pulses required increases with
the dimension $d$ but it may not be a problem in experiments when $d$ is not large. We hope this work
will stimulate experimental activities in the near future.

\begin{center}
\textbf{ACKNOWLEDGEMENTS}
\end{center}

C.P. Yang and Q.P. Su were supported in part by the Ministry of Science and
Technology of China under Grant No. 2016YFA0301802, the National Natural
Science Foundation of China under Grant Nos 11074062, 11504075, and
11374083, and the Zhejiang Natural Science Foundation under Grant No.
LZ13A040002. J.M. Liu was supported in part by the National Natural Science
Foundation of China under Grant Nos 11174081 and 11134003, the National
Basic Research Program of China under Grant No. 2012CB821302, and the
Natural Science Foundation of Shanghai under Grant No. 16ZR1448300. This
work was also supported by the funds from Hangzhou City for the
Hangzhou-City Quantum Information and Quantum Optics Innovation Research
Team.

\end{document}